\documentclass[a4paper,fleqn]{cas-dc}

\usepackage[authoryear,longnamesfirst]{natbib}
\usepackage{subcaption}
\usepackage[normalem]{ulem}
\usepackage{xcolor}
\def\tsc#1{\csdef{#1}{\textsc{\lowercase{#1}}\xspace}}
\tsc{WGM}
\tsc{QE}

\makeatletter
\def\ps@pprintTitle{%
 \let\@oddhead\@empty
 \let\@evenhead\@empty
 \def\@oddfoot{\footnotesize\itshape Preprint submitted to Computational Materials Science\hfill\today}
 \let\@evenfoot\@oddfoot}
\makeatother

\begin{document}
\let\WriteBookmarks\relax
\def\floatpagepagefraction{1}
\def\textpagefraction{.001}

\shorttitle{AI estimation of partition functions with the PULSE method for the thermodynamic properties of disordered compounds}    

\shortauthors{B. Bernard, L. Messina, E. Kawasaki, E. Bourasseau}  

\title [mode = title]{AI-driven estimation of partition functions with the PULSE method: application to the Ising model as a benchmark for the thermodynamic properties of chemically disordered compounds}  



%

\author[inst1]{Baptiste Bernard}[orcid=0009-0004-2391-4726]
\author[inst1]{Luca Messina}[orcid =0000-0003-0562-9070]
\cormark[1]
\cortext[1]{Corresponding author}
\author[inst2]{Eiji Kawasaki}[orcid=0000-0001-7434-4744]
\author[inst1]{Emeric Bourasseau}[orcid=0000-0003-0211-6923]

\affiliation[inst1]{CEA, DES, IRESNE, DEC, Cadarache, F-13108 Saint-Paul-Lez-Durance, France}
\affiliation[inst2]{Universite Paris-Saclay, CEA, LIST, F-91120, Palaiseau, France}









\begin{abstract}
Predicting the thermodynamic properties of chemically disordered materials is constrained by the combinatorial growth of possible atomic configurations, which makes conventional sampling approaches computationally costly. In this work, we extend the PULSE (Partition function Unsupervised Learning Sampling and Evaluation) approach that enables the estimation of observable-dependent partition functions, thereby extending its applicability from local atomic quantities to general thermodynamic observables. This is based on a generative machine-learning approach in which a single model both generates statistically representative atomic configurations and provides explicit partition function estimates. Our primary goal is to assess it on the two-dimensional Ising model, an exactly solvable benchmark system sharing the same statistical-mechanical formulation as chemically disordered binary materials, and compare its performance against exact brute-force calculations as well as Wolff and Wang-Landau Monte Carlo algorithms. PULSE reproduces the relevant partition functions accurately using a limited number of configurations, at a sampling cost that does not grow with system size. These key results demonstrate the relevance of our method to enable the study of disordered materials at scales that are difficult to achieve using conventional approaches, and with limited computational effort.
\end{abstract}




\begin{keywords}
Chemical disorder\\
Machine Learning\\
Partition function\\
Ising\\
Atomic scale\\
 \sep \sep \sep
\end{keywords}

\maketitle

\section{\label{sec:Introduction}Introduction}

Chemically disordered multicomponent materials, such as high-entropy alloys (HEA) \cite{wei2018metastability, tsai2014high} and mixed oxides (MOX), are of growing technological importance for energy and nuclear materials \cite{OSTOVARIMOGHADDAM2021105504, e15104504, pickering2021high, ZINKLE2017569, zhang2008solid}. However, the accurate prediction of their thermodynamic properties by atomic-scale modelling is difficult, as the random occupation of lattice sites by different atomic species leads to an exponential growth in the number of possible configurations. Computing these properties therefore requires sampling this configuration space, which rapidly becomes intractable as the system size and the number of components increase.

Although traditional methods exist for sampling and computing properties for these materials \cite{shin2006thermodynamic, vigier2015structural}, they have inherent flaws that make them ineffective in certain respects. For instance, the use of Special Quasirandom Structures (SQS) is a common approach to study high-entropy alloys \cite{Gao2016, case2016convergence, wei1990electronic}, but lacks criteria for determining the representativeness of the generated configurations. Also, classic Monte Carlo methods, although capable of estimating target properties, are computationally inefficient as they require the generation of a large number of configurations before convergence, making them very costly. For instance, a Monte Carlo evaluation of thermodynamic properties in MOX fuels needed over 100 million configurations to converge  \cite{TAKOUKAMTAKOUNDJOU2020152125}. 

Recently, the approach to this type of problem has been strongly impacted by the introduction of Machine Learning (ML) methods, which make possible to optimize overly complex calculations thanks to their generalization capabilities. This is especially true since the advent of deep learning \cite{lecun2015deep} and neural networks applied to the study of physical systems. This type of approach has proven effective when modelling disordered physical systems \cite{liu2023machine}, in particular by introducing innovative ways of addressing the atomic scale in a multiscale modelling  framework \cite{TAN2023100114} using unsupervised ML methods. In this context, we have introduced in a previous work PULSE \cite{karcz2024targeting} (Partition function Unsupervised Learning Sampling and Evaluation for disordered compounds), a general ML tool that aims to estimate observable properties of an atomic-scale physical system formulated as partition functions. Inspired by an inverse variational autoencoder architecture (VAE) \cite{cantwell2022approximate}, this method uses generative sampling to find the smallest set of significant configurations of the system in order to compute its mean properties. However, this tool was designed to estimate local properties such as defect formation energies in the atomic lattice \cite{karcz2024targeting}, and was therefore not suitable for obtaining thermodynamic properties, which are inherently global and thus need considering atomic systems of much larger sizes. 

In this work, we develop the theoretical and computational framework that allows the PULSE method to estimate general observable properties of multi-component systems, including thermodynamic properties, formulated as ratios of partition functions. This is an important step beyond our previous work \cite{karcz2024targeting} that targeted local properties only. In this context, the intent is to conduct low-cost experiments while improving the method in a concrete and relevant way for future applications. The method is self-supervised and data-driven: a single generative model learns to produce statistically representative configurations, and retuns an explicit estimate of the partition function. Most importantly, the goal is to minimise the number of configurations needed,  thereby minimising the computational resources required to estimate a property.  As a first application, we focus in this work on a model system governed by equations similar to those of the target materials. The well-known Ising system  appears to be an ideal testing ground for our target approach. Indeed, this system has been extensively studied from a sampling perspective \cite{landau1976finite, wolff1989collective} because it is analytically solvable in the 2D case without an external magnetic field \cite{onsager1944crystal} and because calculations on this system can be carried out at low cost. This allows us to benchmark the results of PULSE against exact, brute-force solutions and traditional Monte Carlo methods. . Moreover, over the last ten years, a wide range of ML models have been exploring the properties of this system, ranging from supervised \cite{zhang2021ising, torlai2016learning} to unsupervised learning \cite{d2020learning, alexandrou2020critical, wetzel2017unsupervised, hu2017discovering, walker2020deep, alamino2024explaining, yevick2022variational}. We thus use the Ising model as an analogue of a two-component chemically disordered system, and we probe its physical properties with PULSE.

After introducing the extended PULSE mathematical framework and the Ising counterpart, we conduct performance tests on our method for estimating magnetic susceptibility. We continue this testing phase by investigating scaling as we increase the size of the Ising systems under study. Finally, we compare our method with other Monte Carlo methods suited to the study of the Ising system. 

Once the results are presented, we draw conclusions that will show us how to apply PULSE on the prediction of thermal properties for more complex multi-component systems, such as MOX and HEA for nuclear applications.

\section{\label{sec:Methodo}PULSE methodology}

In this section, we introduce the main features of PULSE. The principle is to identify and generate the smallest set of significant configurations that provides an accurate estimate of the target property in the large configuration space of a chemically disordered compound. The tool learns the physics of the system through its partition function. 
Sampling seeks the right balance between a configuration's probability (i.e., with low energy) and its contribution to the target property, as there can be low-probability configurations where this property has a large value, and vice versa high-probability ones where the value of this property is small.

The PULSE method \cite{karcz2024targeting} is based on unsupervised VAE \cite{kingma2013auto} ML models. However, VAEs considered here are used in an "inverse" version, in order to reproduce the hidden complex distribution (i.e., that of atomic configurations) by generating samples from a chosen and simple distribution. With respect to classic VAEs, the role of encoder and decoder is thus inversed.

PULSE training proceeds via an unsupervised loop with no prior dataset. In each iteration, PULSE samples configurations from a latent space.  The model is trained in such a way that, for each iteration, configuration energies are computed  with a dedicated method of choice (for instance, \textit{ab initio} calculations, or interatomic potentials). The calculated energies are used to determine the error, which enables backpropagation to update the weights of the network and the latent space. This process is repeated until convergence and then the value of the loss provides an estimate of the target partition function.

The goal of the present study is to improve the PULSE method by generalizing the approach to any partition function defined with respect to an observable property of the system. In what follows, we introduce a general formulation that allows us to describe a generic thermodynamic observable property.

\subsection{\label{sec:Form}Formulation of the target property in the PULSE framework}

Let $O$ be an observable property of a given physical system whose possible discrete states $x$ belong to the configuration space $X$. We define the partition function of this property at temperature $T$ by $Z_{O}(T)$ :

\begin{equation}
    Z_{O} (T) = \sum_{x \in X} O(x) \exp\left(\frac{-E(x)}{k_{B}T}\right)  \; .
\end{equation}

A way to define the thermodynamic average of the property among the possible configurations is:

\begin{equation}
    \langle O (T)\rangle = \frac{Z_{O}(T)}{Z(T)} \; .
\end{equation}
with $Z(T)$ being the canonical partition function of the physical system. Our objective is to calculate this average by estimating both partition functions independently using the PULSE algorithm.

To do so, we define $P_{O}(x)$ as the probability for the property to have the value $O(x)$ in the configuration space for a given temperature $T$ and configuration $x$ (a brief demonstration is provided in Appendix A):
\begin{equation}
    P_{O}(x) = \frac{\exp\left(\frac{-E(x)}{k_{B}T} + \ln(O(x)) \right)}{Z_{O}(T)} \; .
    \label{eq:pO_probability}
\end{equation}

This probability $P_{O}(x)$, targeted by PULSE, is a way to reconstruct $\langle O (T)\rangle$. Since $P_{O}(x)$ contains the logarithm of the property $O(x)$, this property must be strictly positive for each sample. Therefore, given that the observable can take various forms, it is necessary to ensure positivity of $O(x)$ in order for Eq. \ref{eq:pO_probability} to hold. We use a suited function built in such a way we can use it to sample a wide range of values of the observable despite the logarithm shape. Examples of this procedure are provided in the following sections.

Ensuring positivity therefore constitutes the fundamental and minimal theoretical requirement imposed by the generalized formulation. The extension beyond local properties does not result from the positivity constraint itself, but from the generalization of the framework to arbitrary observables, which are no longer required to be defined locally. Within this generalized formulation, positivity is the only remaining constraint on the target observable, allowing the approach to be applied to both local and global properties.

\subsection{\label{sec:Learing}Learning process}

In this section, the goal is to learn $P_{O}(x)$ for each $T$. Since the space in which $x$ lies is too large for a systematic exploration, we use an approach that involves constructing and using another distribution $P(y)$ that can be easily sampled from in order to recover $P_{O}(x)$. The auxiliary variable $y$ acts as a compressed lower-dimensional representation of $x$; its dimensionality is thus significantly smaller than that of $x$. To link these two distributions, we define a probability distribution $R(x|y)$. Since this distribution is arbitrary, its form must be optimised so that sampling $x$ from it reproduces the target distribution $P_{O}(x)$.

To link $R(x|y)$ to $P_{O}(x)$, we then use Bayes' theorem with another conditional distribution $Q(y|x)$ that allows for inverse sampling:
\begin{equation}
    P_{O}(x) Q(y|x) = P(y) R(x|y) \; .
    \label{eq:Bayes}
\end{equation}

We then transform the ideal scenario of Eq. \eqref{eq:Bayes} into an optimisation problem, in which distributions $R$ and $Q$ (given in Appendix B, Eqs. B1-B3) are parameterised so that samples $x$ drawn from $P(y)$ follow the target distribution $P_{O}(x)$. To do so, we use a classic tool to measure the difference between two distributions, \textit{i.e.}, the Kullback-Leibler (KL) divergence \cite{kullback1951information} defined as follows:

\begin{equation}
\begin{split}
D_{KL}\left[P(y) R(x|y)  \parallel P_{O}(x) Q(y|x)\right]\\
= \mathbb{E}_{\,y \sim P, \,x \sim R} \, \ln \left[ \frac{P(y) R(x|y)}{P_{O}(x) Q(y|x)} \right]\\
= -\mathbb{E}_{\,y \sim P, \,x \sim R} \, \ln \left[ \frac{P_{O}(x) Q(y|x)}{P(y) R(x|y)} \right] \; .
\end{split}
\end{equation}

Since the KL divergence is always non-negative, we can write:
 
\begin{equation}
 0 \leq -\mathbb{E}_{\,y \sim P, \,x \sim R} \, \ln \left[ \frac{P_{O}(x) Q(y|x)}{P(y) R(x|y)} \right] \; .
\end{equation}
and finally:

\begin{equation}
 \ln Z_{O} \geq \mathbb{E}_{\,y \sim P, \,x \sim R} \, \ln \left[ \frac{\exp\left(\frac{-E(x)}{k_{B}T} + \ln(F(O(x))) \right) Q(y|x)}{P(y) R(x|y)} \right] \; .
 \label{eq:elbo}
\end{equation}

\begin{figure}
\includegraphics[width=\linewidth]{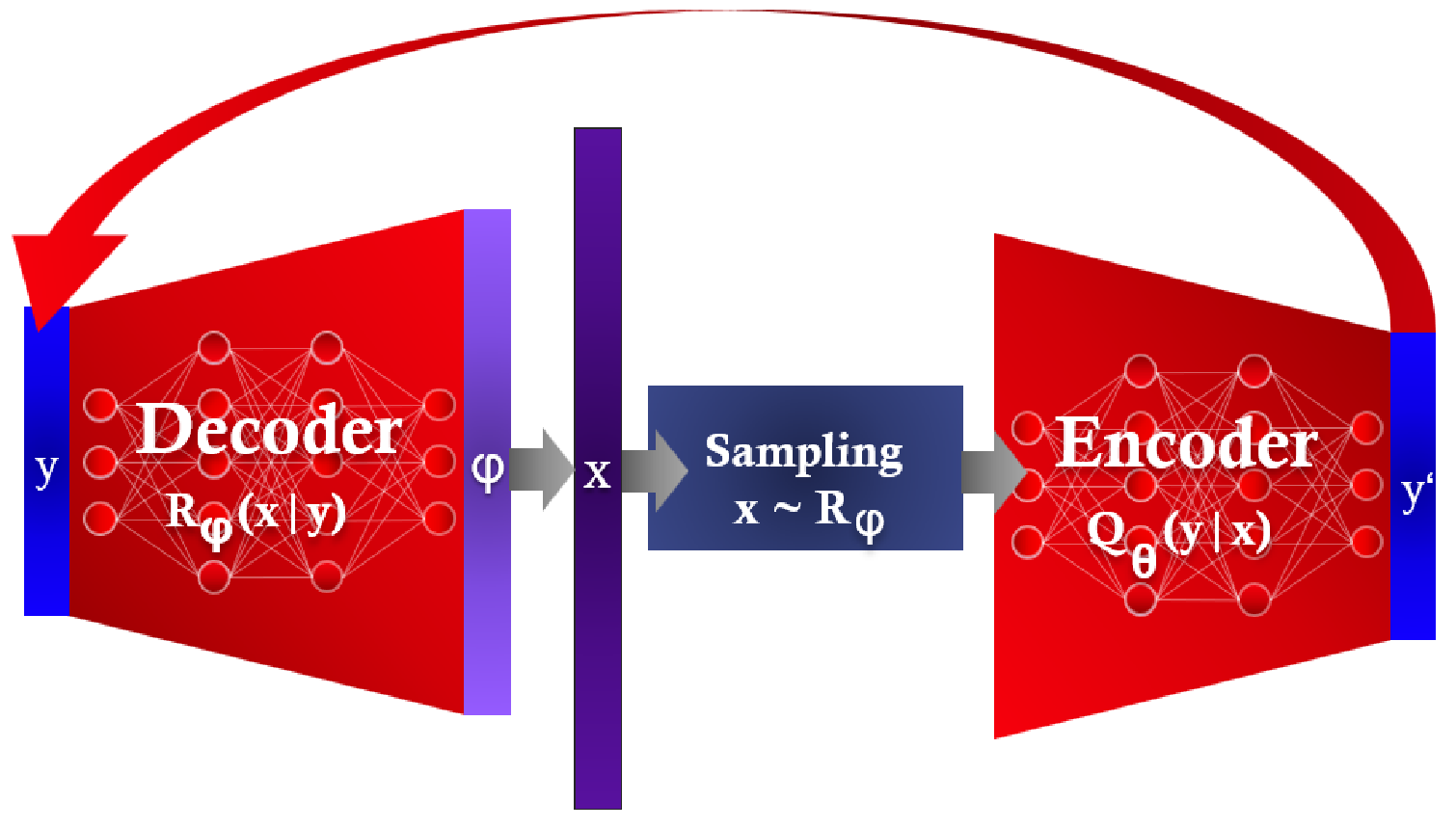}
\caption{\label{fig:PULSE}Schematic view of the PULSE inverse VAE architecture.}
\end{figure}

Hence, by maximizing the right-hand side of Eq.  \eqref{eq:elbo}, we ensure that our model provides a lower bound of the partition function $Z_{O}$. In practice, our goal is to saturate this inequality and provide a reliable estimate of the partition function. More specifically, as additional configurations are generated and incorporated into the training procedure, the evidence lower bound (ELBO) approaches its asymptotic value. Learning thus involves solving this optimization problem, based on which it is possible to build an ML architecture that is suitable for solving this type of problem. The parameters of the joint distributions $R$ and $Q$ are set via two neural networks (a decoder and an encoder, respectively) that, as learning converges, will enable the ideal equality of \eqref{eq:Bayes} to be approximated. Figure \ref{fig:PULSE} shows the inverse-VAE architecture of our learning model: the networks encode and decode the information at the edges of the architecture in the latent space $y$. In each iteration, the model samples $y$ from $P(y)$ and then samples configurations of the system $x$ from $R(x|y)$. Since $x$ is discrete, the Gumbel softmax trick \cite{jang2016categorical} is applied to generate the samples from this continuous distribution. After calculations of energies and observable properties, results are given back to reconstruct a new latent space $y'$ from $Q(y|x)$.  

All PULSE networks use the same architecture shown in Fig. \ref{fig:PULSE}, where the encoder and the 
decoder have the same architecture: three layers with a single hidden layer containing 2\,048 nodes, with SELU activation functions. Learning proceeds using the Adam optimizer \cite{kingma2014adam} and a learning rate of $10^{-4}$.  The temperature parameter of the Gumbel softmax function \cite{jang2016categorical} is set to $\tau = 1/10$. 

\clearpage

\section{\label{sec:Ising}Ising model}

In this section, we introduce the Ising model and the properties chosen to assess our method. The objective is twofold: first, to review the equations that govern this system, which will form the mathematical framework for our analysis; second, to establish an analogy between this simplified physical model and the systems targeted by PULSE.  In this context, the Ising model serves as a well-understood version of a disordered two-component system that allows us to conduct an initial assessment of our method applied to the prediction of global thermodynamic properties. It should be mentioned that the 2D Ising system routinely serves as a benchmark for several developments in machine learning methods \cite{d2020learning, alexandrou2020critical, wetzel2017unsupervised, hu2017discovering, walker2020deep, alamino2024explaining}, which have proved their effectiveness when applied to this system. It consists of a square lattice of size $N = L\times L$, where each site is occupied by a spin.
These spins $S_i$ can either be 'up' or 'down', i.e., $S_i = \pm 1 \; \forall i \in \{1,\dots,N\}$).

Interactions between these spins are described according to the following expression for the total energy $E$ of the system, where $i,j$ denotes nearest-neighbor interaction only: 

\begin{equation}
    E =  - J \sum_{i,j}S_{i}S_{j} -  H \sum_{i=1}^{N} S_{i} \; .
\end{equation}

We consider the case with no external magnetic field ($H = 0$). The exchange energy $J$, controlling the interaction strength, is set to $J=1$, which corresponds to the ferromagnetic case. We define the magnetization as the sum of the spins over the system: 

\begin{equation}
    M = \sum_{i=1}^{N} S_{i} \; .
\end{equation}

In this configuration, the Ising system has two regimes that depend on temperature. At high temperature, the system is disordered, with no preferred spin orientation. At low temperature, spins are aligned either up or down, since there is no external magnetic field,  with a phase transition occurring at the Curie temperature $T_\mathrm{C} = 2/$$\ln(1+\sqrt{2})$ $\approx 2.2693$K. 

Our central interest in the study of this model lies in the analysis of the physical quantities of interest such as:

\begin{itemize}
    \item[-] The susceptibility $\chi$:
\begin{equation}
    \chi= \frac{\partial M}{\partial H} = N \beta  (\langle M^{2}  \rangle- \langle M \rangle^{2}) \; .
    \label{eq:susdef}
\end{equation}

    \item[-] The specific heat at constant volume $C_\mathrm{V}$:
\begin{equation}
    C_\mathrm{V}= \frac{\partial E}{\partial T} = N \beta^{2}  (\langle E^{2} \rangle - \langle E \rangle^{2}) \; .
    \label{eq:cvdef}
\end{equation}
with $\beta=\frac{1}{k_\mathrm{B}T}$ and $k_\mathrm{B} = 1$ for convenience.\\
\end{itemize}

For the latter property, we can directly relate this formulation for a 2D Ising system to the one that can be used for the heat capacity at constant volume in real physical compounds \cite{lagache2001prediction}:

\begin{equation}
    C_\mathrm{V} =
    \frac{1}{k_{B} T^{2}} \left(\langle U^{2} \rangle - \langle U \rangle^{2} \right) \; .
\end{equation}

where $U$ represents the internal energy of the thermodynamic system. Indeed, this property is formulated as a combination of observable average properties of a chemically disordered system, so it can be sampled and estimated in the same way as the properties mentioned above in the Ising model framework.

We can write the thermodynamic averages of these quantities as ratios between different partition functions. For instance, for a given temperature $T$, let $x$ be a given spin configuration. The average of the total magnetic moment is:

\begin{equation}
    \langle M (T)\rangle = \frac{\sum_{x \in X} M_{x}(T)\exp\left(-\beta E_{x}(T)\right)}{\sum_{x \in X}\exp\left(-\beta E_{x}(T)\right)} \; .
\end{equation}

We thus define two partition functions as follows:

\begin{equation}
    Z(T) = \sum_{x \in X}\exp\left(- \beta E_{x}(T)\right) \; .
    \label{eq:can}
\end{equation}

\begin{equation}
    Z_{M}(T) = \sum_{x \in X} M_{x}(T) \exp\left(-\beta E_{x}(T)\right) \; .
\end{equation}
\\
so that:

\begin{equation}
    \langle M (T)\rangle = \frac{Z_{M}(T)}{Z(T)} \; .
    \label{eq:M}
\end{equation}

However, $\langle M\rangle$ cannot be calculated as such with PULSE because this quantity can be either positive or negative, while PULSE must learn on strictly positive values. Thus, in this work, we decide to target $\langle|M|\rangle$ instead of $\langle M\rangle$, which transforms $\chi$ into a distinct magnetic susceptibility $\chi'$. This susceptibility $\chi'$ is still a quantity commonly studied in the Ising system \cite{murtazaev2015critical, albano2003corner, shchur2001critical}. We thus have:

\begin{equation}
     Z_{|M|}(T) =  \sum_{x \in X} \exp\left(-\beta E_{x}(T) + \ln|M_{x}(T)|\right) \; .
     \label{eq:mag}
\end{equation}

\begin{equation}
    Z_{M^{2}}(T) =  \sum_{x \in X} \exp\left(-\beta E_{x}(T) + \ln M_{x}^{2}(T)\right) \; .
    \label{eq:magsquare}
\end{equation}

To determine the susceptibility $\chi'$, we rewrite equation \eqref{eq:susdef} as:

\begin{equation}
    \chi'(T) = N \beta  \left(\frac{Z_{M^{2}}(T)}{Z(T)} - \left(\frac{Z_{|M|}(T)}{Z(T)}\right)^{2} \right) \; .
    \label{eq:sus}
\end{equation}

The same development can be made for the specific heat at constant volume written as follows :

\begin{equation}
    C_\mathrm{V}(T) = N \beta^{2}  \left(\frac{Z_{E^{2}}(T)}{Z(T)} - \left(\frac{Z_{E}(T)}{Z(T)}\right)^{2} \right) \; .
    \label{eq:spheat}
\end{equation}

The associated partition functions are then :

\begin{equation}
     Z_{E}(T) =  \sum_{x \in X} \exp\left(-\beta E_{x}(T) + \ln [-E_{x}(T)] \right) \; .
     \label{eq:energy}
\end{equation}

\begin{equation}
    Z_{E^{2}}(T) =  \sum_{x \in X} \exp\left(-\beta E_{x}(T) + \ln \left[E_{x}^{2}(T)\right]\right) \; .
\end{equation}

In Eq. \eqref{eq:energy}, we sample $- E$ instead of $E$ to calculate the associated partition function in Eq. \eqref{eq:energy}. To deal with cases where the energy is positive – which are extreme cases, given that the average energy of this system is negative – we seek to exclude them from the training process by treating them as a near-zero ($< 10^{-8}$) observable. This choice is physically justified because very high-energy configurations do not contribute to the partition function especially in the low-temperature regime, where these high-energy values have low probability. As our sampling aims to calculate the partition function, the error made in excluding high-energy configurations should remain negligible.

\section{\mbox{Results}}

In this section, we assess the robustness of our method using several use cases and benchmarks. The first part examines the reliability of our model in estimating the aforementioned Ising properties. The second part demonstrates its scalability with respect to system size. Finally, we compare our method to specific Monte Carlo algorithms to evaluate the relative performance of our method.

As shown in Fig. \ref{fig:PULSE}, the PULSE model trains two networks, an encoder and a decoder, for each of the partition functions and for each temperature needed to reconstruct the thermodynamic data mentioned in Sec. \ref{sec:Ising} :  $Z$, $Z_{|M|}$, $Z_{M^{2}}$, $Z_{E}$, and $Z_{E^{2}}$. Exact values are obtained by "brute-force" computations where all possible spin configurations are accounted for.

\subsection{ Benchmark on Ising properties}

In this subsection, we use PULSE to estimate the specific magnetic susceptibility $\chi'$ of the Ising system. Results are normalised by the total number of spins $N$ in the system. We apply this study to a 4$\times$4 Ising lattice, which has 65\,536 different configurations.

\begin{figure}
    \centering
    \includegraphics[width=\linewidth]{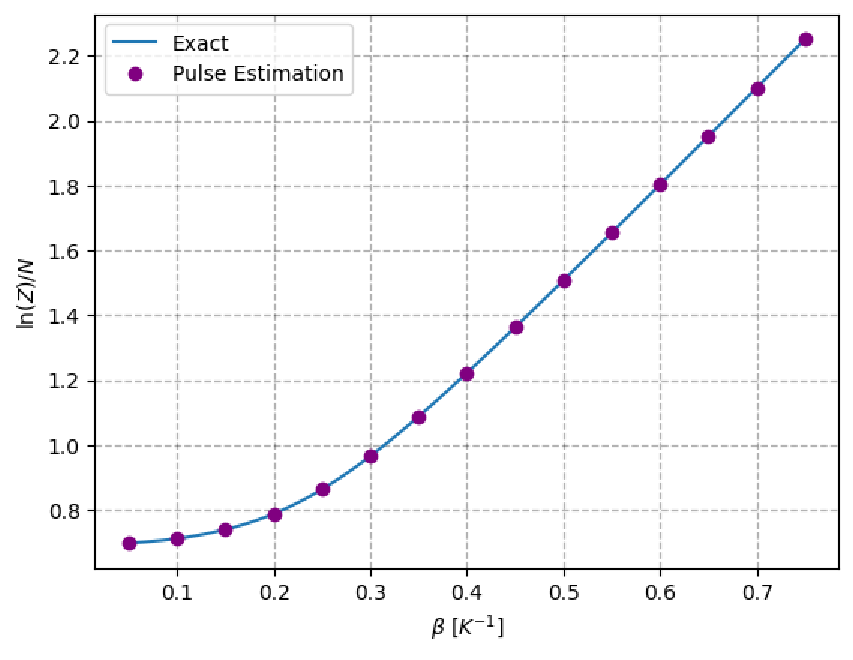}
    \caption{Canonical partition function of the $4 \times 4$ Ising system, with the exact value (obtained by including all possible spin configurations in the calculation) shown as a blue curve and the PULSE estimation as purple dots. The stopping criterion is set to  a 1\% error relative to the exact value.}
    \label{fig:Zcanonic}
\end{figure}

First, we estimate the partition functions necessary to build the target property as defined in Eq. \eqref{eq:sus}. In Fig. \ref{fig:Zcanonic}, we show the PULSE estimate of the canonical partition function of the system (Eq. $\eqref{eq:can}$), in good agreement with the exact solution, confirming the results already achieved in a previous work \cite{cantwell2022approximate} where the inverse VAE concept was introduced and applied to the 2D Ising partition function. Here, PULSE reaches a remarkable accuracy, with an error below 1\% compared to the exact value, using between 200 and 2\,500 samples per temperature. The 1\% criterion is not the convergence criterion intrinsically used by the PULSE method. It is primarily used in the present study as an external validation criterion, made possible by the availability of the exact solution for the 2D Ising model.

\begin{figure}
    \centering
    \includegraphics[width=\linewidth]{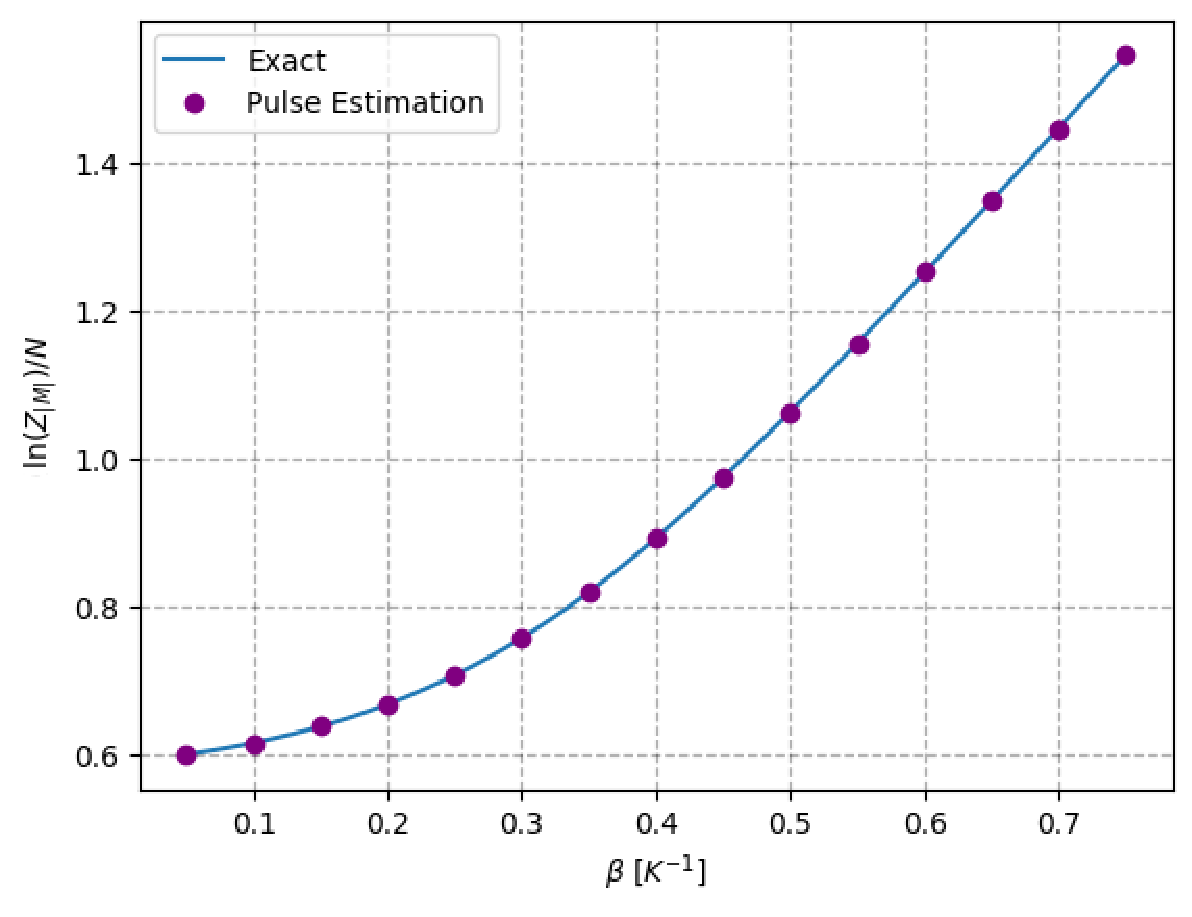}
    \caption{Partition function of the absolute value of magnetization for the $4 \times 4$ Ising system, with the exact value (obtained by including all possible spin configurations in the calculation) shown as a blue curve and the PULSE estimation as purple dots. The stopping criterion is set to a 1\% error relative to the exact value.}
    \label{fig:Zmag}
\end{figure}

Figure \ref{fig:Zmag} represents the PULSE estimation of $Z_{|M|}$, \textit{i.e.}, the partition function associated with the absolute value of the magnetization (Eq. \eqref{eq:mag}). Similarly to the previous quantity, PULSE reaches an error below 1\% compared to the exact value, using between 200 and 2\,500 samples per temperature, depending on the target temperature. This shows that PULSE consistently achieves an outstanding accuracy on the partition function associated with specific properties, using a relatively low and stable amount of configurations. 

This high precision in our estimations allows us to properly predict properties that depend on those partition functions, as shown in Fig. \ref{fig:mag,magsquare,sus} for the properties needed to obtain the modified susceptibility $\chi'$. These plots show respectively estimations of average modified magnetization, squared magnetization, and modified susceptibility as presented in Eqs. \eqref{eq:M} through \eqref{eq:sus}. Purple symbols correspond to quantities obtained by combining the different estimations of PULSE at each temperature.

For properties that rely on the combination of two partition functions, \textit{i.e.}, equal to the ratio between the partition function of the observable property and the canonical partition function, as in Eq. \eqref{eq:M}, the PULSE evaluation is in excellent agreement with the exact solution (see  Figure \ref{fig:mag,magsquare,sus} (a) for the average modified magnetization, and Figure \ref{fig:mag,magsquare,sus} (b) for the average squared magnetization). On the other hand, properties that are combination of more than two partition functions, as in Eq. \eqref{eq:sus}, present slight deviations at low temperature. This is the case for the modified magnetic susceptibility (see  Figure \ref{fig:mag,magsquare,sus} (c)). 

\begin{figure}

    \begin{subfigure}[b]{0.91\linewidth}
        \caption{}
        \includegraphics[width=0.91\linewidth]{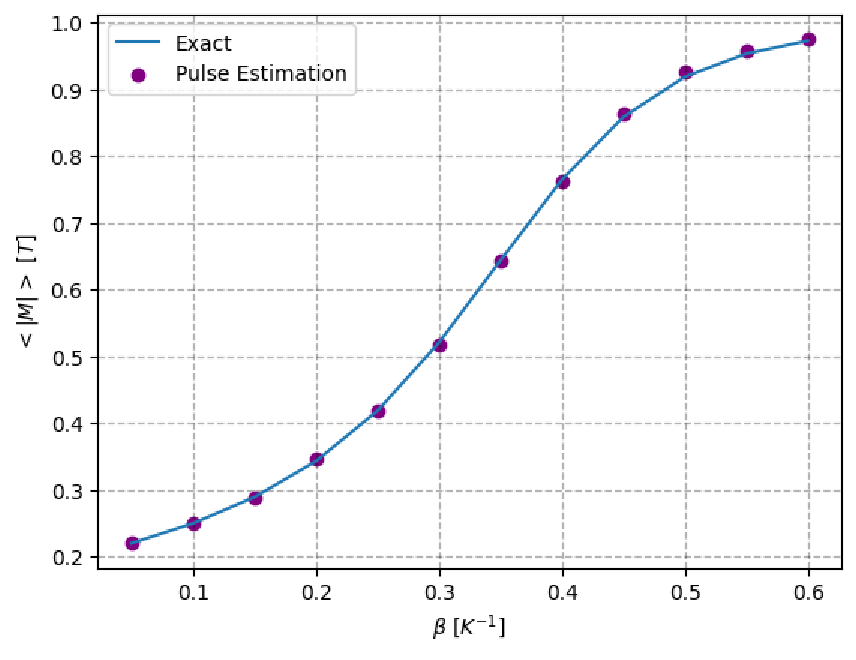}
    \end{subfigure}
    \begin{subfigure}[b]{0.91\linewidth}
        \caption{}
        \includegraphics[width=0.91\linewidth]{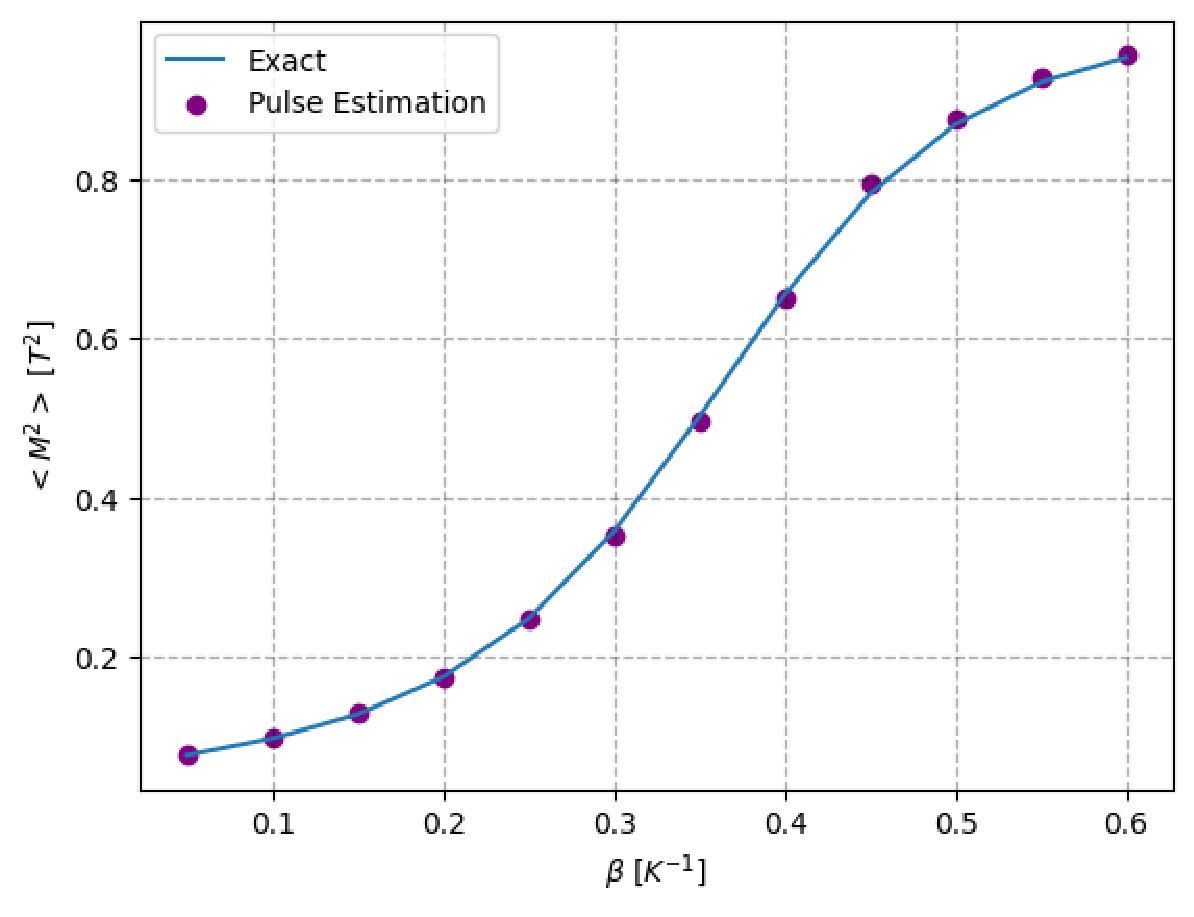}
    \end{subfigure}
    \begin{subfigure}[b]{0.91\linewidth}
        \caption{}
        \includegraphics[width=0.91\linewidth]{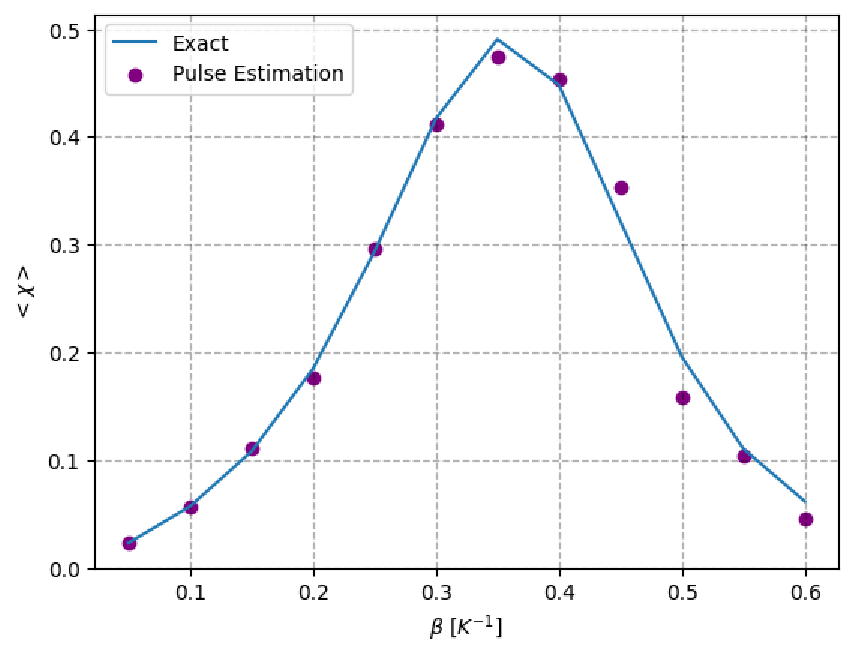}
    \end{subfigure}
    \caption{Average values of (a) normalized magnetization, (b) squared magnetization, and (c) modified susceptibility obtained as recombinations of partition functions sampled by PULSE. Purple dots are recombinations of partition functions according to equations \eqref{eq:can}, \eqref{eq:mag}, and \eqref{eq:magsquare}. The blue curve marks the exact value (obtained by including all possible spin configurations in the calculation). }
    \label{fig:mag,magsquare,sus}
\end{figure}

This likely arises because at low temperature (and therefore high $\beta$), the Ising system is restricted to two significant states that correspond to the configurations where all spins are aligned either 'up' or all 'down'. PULSE has difficulties finding these states and performs better at high temperatures, when more configurations contribute significantly to the partition function. In addition, the errors in the various estimates accumulate in the reconstruction of the property. Nevertheless, it is important to note that the results are displayed in Fig. \ref{fig:mag,magsquare,sus} on a linear scale, which accentuates errors that are otherwise negligible on the logarithmic scale on which our model operates. For instance, for $\beta = 0.5$ K$^{-1}$, the error is $15\%$ on a linear scale, but only $2.7\%$ on a logarithmic scale. The performance of the method is therefore satisfactory: accurate predictions of the target properties can be obtained with few samples and thus low computational cost.

\bigskip
\bigskip
\bigskip
\bigskip
\bigskip
\bigskip
\bigskip
\bigskip
\bigskip
\bigskip
\bigskip
\bigskip


\subsection{Scalability}

\begin{figure}
    \centering
    \includegraphics[width=\linewidth]{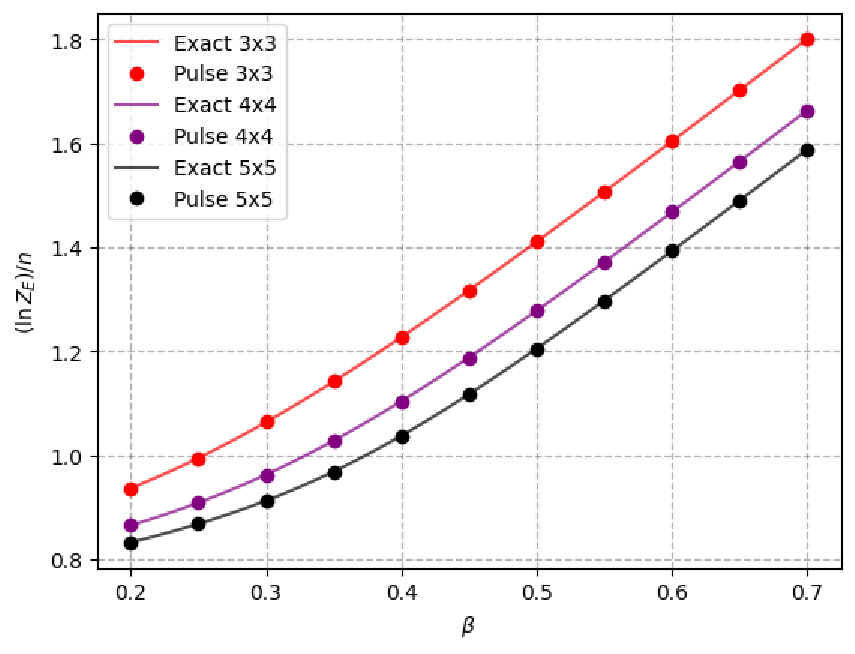}
    \caption{Energy-related partition functions of 2D Ising systems sampled by PULSE for different system sizes : $3\times3$ (512 configurations), $4\times4$ (65\,536 configurations) and $5\times5$ (33\,554\,432 configurations), compared to exact values (obtained by including all possible spin configurations in the calculation).}
    \label{fig:Zsize}
\end{figure}

In this section, we investigate the robustness of our sampling method with respect to system size to assess its stability and scalability. In Fig. \ref{fig:Zsize}, we compare three different evaluations of the partition function associated with energy $Z_E$, as in Eq. \eqref{eq:energy}, to the exact values for Ising systems of size varying from 3$\times$3 to 5$\times$5. For these calculations, we consider the learning converged when the estimated error is below 1\%.

For any system size, the amount of configurations needed in the sampling procedure does not exceed 2\,000 samples per temperature, and we find that the larger the system, the fewer samples are actually needed. For a 3$\times$3 Ising system, the algorithm requires on average between 1\,500 and 2\,000 samples, whereas for a system of size 5$\times$5, it requires around 500 to 1\,000 samples per temperature. 

This result provides initial evidence of the good scalability of our method: system size has a small impact on the amount of configurations needed to obtain a satisfying accuracy, and the algorithm seemingly converges better as the system has a larger number of spins. Furthermore, this result also allows us to confirm the validity of our method for sampling $Z_{E}$, despite the neglect of states with positive energy discussed in Section \ref{sec:Ising}. 

The fact that, for the $3\times3$ Ising system, our method requires more samples than available states in the configuration space ($2^{(3\times3)}=512$) can be explained by the fact that our method is not designed to handle such small systems, where exhaustive exploration is feasible. This architecture is intended to infer properties of very large configuration spaces, as we will show below.

Because of the exponentially increasing number of possible configurations, it is difficult to compare with the exact solution at larger scales. From the $6\times6$ system onwards, there are several billion configurations to sample to calculate the exact solution (over $6.8 \times 10^{10}$ samples for $6\times6$ and over $5.62 \times 10^{14}$ samples for $7\times7$). 

In Fig. 6, we demonstrate the scalability of our method by extending it to larger Ising system sizes (from 3$\times$3 to 60$\times$60). Each learning process is performed with 2\,000 samples using the same PULSE model architecture. The objective is to examine the trends in the curves produced by PULSE and assess if, despite the increase in system size, the expected behavior is recovered. Each curve is obtained from 10 independent models in order to average the estimations and provide error bars. 

\begin{figure}
    \centering
    \includegraphics[width=\linewidth]{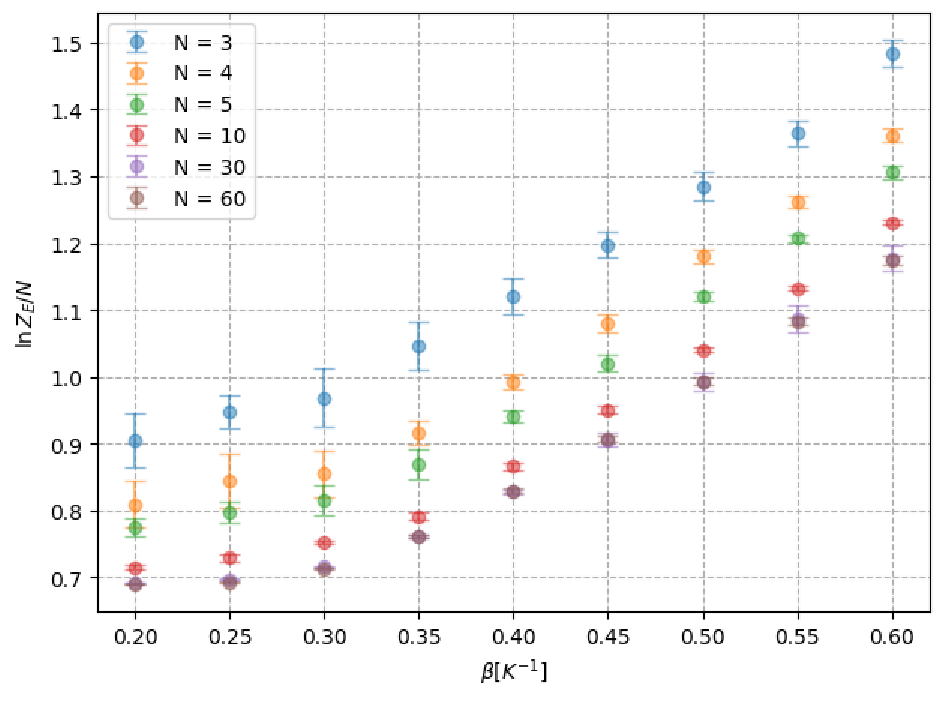}
    \caption{Energy-related partition functions of 2D Ising systems estimated by PULSE for system sizes ranging from 3$\times$3 to 60$\times$60.}
    \label{fig:Zsizehd}
\end{figure}

We can see that the trend estimated in Fig. \ref{fig:Zsize} is accurately reproduced: ln Z$_{E}$/N increases with increasing $\beta$ and decreases. We also find that convergence improves as the number of spin increases: indeed, for all temperatures, we observe a reduction in error bars at large system sizes. 

In light of these results, we conclude that PULSE is robust with respect to system size: without changing the size of the model, very few samples (relative to the size of the configuration space) are needed to achieve convergence. This is a clear advantage for studying real, more complex physical systems. 

\subsection{Comparison with Monte-Carlo estimates}

In this last section, we compare our results with another typically method used to estimate mean properties in disordered compounds: Monte Carlo. For sampling the Ising lattice system, we use two Monte Carlo algorithms specifically designed for this type of systems: Wolff \cite{wolff1989collective} and Wang-Landau \cite{wang2001determining}.

Instead of randomly flipping individual spins, as in conventional Monte Carlo methods, Wolff's algorithm flips entire clusters of spins. As a result, correlated spins are flipped simultaneously, allowing for a more efficient exploration of the configuration space. 
However, this method is not designed to estimate partition functions, but rather observable properties of the system. It thus needs to be adapted to enable comparisons with the PULSE method. 

To this end, let $Z$ be the canonical partition function of the Ising system and $|X|$ the number of configurations of this system. We can write:

\begin{equation}
\begin{split}
     \frac{1}{Z} =  \sum_{x \in X} \frac{1}{|X| Z} = \sum_{x \in X} \frac{\exp\left(\beta E(x) - \beta E(x) \right)}{|X| Z} \\
     = \frac{1}{|X|} \sum_{x \in X} \exp(\beta E(x)) p(x) \; ,
\end{split}
\label{eq:wolff1}
\end{equation}
where: 

\begin{equation}
    p(x) = \frac{\exp(-\beta E(x))}{Z} \; .
\end{equation}

So by taking $m$ samples of the system, obtained using the Monte Carlo scheme, we can estimate $Z$ in the following way:

\begin{equation}
\begin{split}
     \frac{1}{Z} =  \frac{1}{|X|} \mathbb{E}_{x \sim p(x)}[\exp(\beta E(x))] \approx \frac{1}{|X| m} \sum_{j=1}^{m} \exp(\beta E(x)) \; ,
\end{split}
\label{eq:wolff2}
\end{equation}
with : 
\begin{equation}
    x \sim p(x) \; .
\end{equation}

The second Monte Carlo method we consider is the the Wang-Landau method, a standard choice for estimating partition functions. It is a Monte Carlo algorithm that, instead of sampling configurations according to Boltzmann's law,  performs a uniform random walk in energy in order to estimate the density of states and ultimately the canonical partition function of the system. Further details on this method can be found elsewhere\cite{wang2001determining}. It should be noted that, since this method must uniformly explore the energy space, it is difficult to ensure convergence for large system sizes. This technique is therefore introduced for comparison purposes, but is not well suited to the thermodynamic property prediction applications considered for PULSE.

\begin{figure}
    \begin{subfigure}[b]{\linewidth}
        \caption{}
        \includegraphics[width=\linewidth]{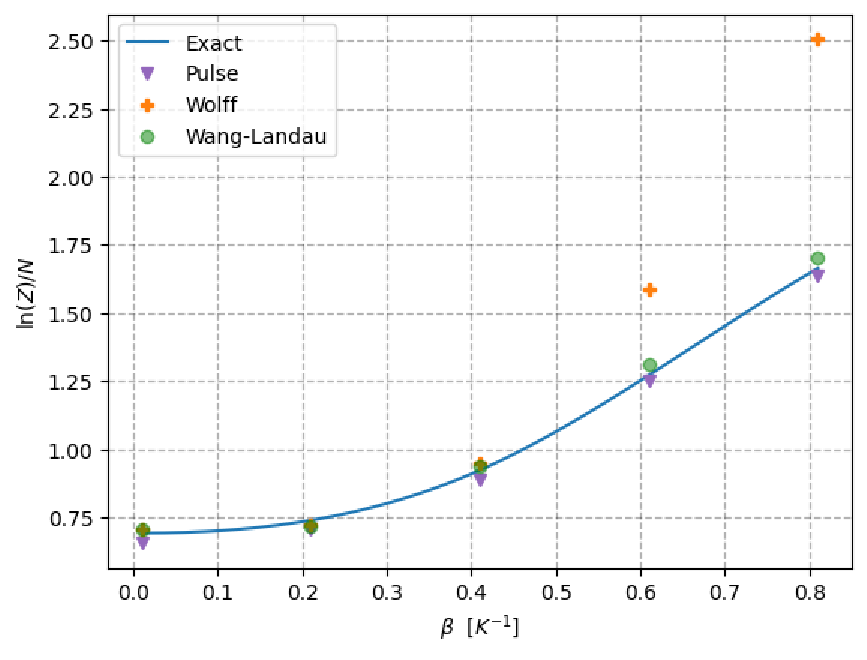}
    \end{subfigure}
    \begin{subfigure}[b]{\linewidth}
        \caption{}
        \includegraphics[width=\linewidth]{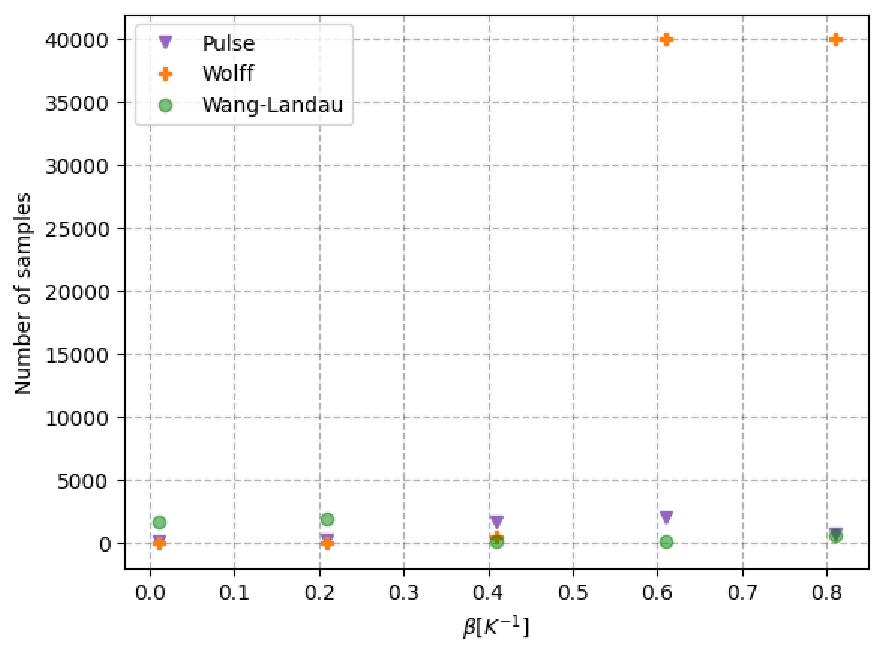}
    \end{subfigure}
    \caption{Comparison of sampling methods for the canonical partition function of the 2D Ising system. (a) Canonical partition function of a $4\times4$ Ising system estimated by different sampling method.  (b) Number of samples required for each sampling at each temperature. }
    \label{fig:McvPULSE}
\end{figure}

We now compare partition function sampling using these methods. In Fig. \ref{fig:McvPULSE} (a), we present results from PULSE, Wolff, and Wang-Landau for the estimation of a canonical partition function, together with the number of samples needed Fig. \ref{fig:McvPULSE} (b). For this comparison, we set the stopping criterion to a 1\% error relative to the exact value. If this precision is not reached after 40\,000 samples, the algorithm is stopped. 

For this partition function, Wolff's algorithm fails to sample the system accurately beyond the phase transition ($\beta \approx$ 0.4 K$^{-1}$). Since convergence is limited to 40\,000 samples by choice, Wolff sampling stops with an error of 29.1\% and 83.9\% for $\beta=0.6$ K$^{-1}$ and $\beta=0.7$ K$^{-1}$, respectively. So, in this context, we see that PULSE provides a significantly better estimate of the canonical partition function of the Ising system, while requiring fewer samples than Wolff, which is designed to estimate thermodynamic properties. This is expected, as Wolff's algorithm does not require explicit evaluation of the partition function to estimate mean properties. Instead, averages are obtained directly from sampled configurations, regardless of whether or not they are representative of the partition function.

Concerning the Wang-Landau algorithm, designed instead to estimate partition functions, results are comparable to those of PULSE, but more configurations are required to converge at high temperatures. Furthermore, results deviate slightly more from the exact value at low temperatures. 

A more relevant comparison is to compare the performance of PULSE and both Monte Carlo methods in estimating thermodynamic properties such as the average energy and the associated heat capacity, as defined in Eq. \eqref{eq:spheat}.

For this comparison, we perform 10 independent evaluations per property, using 2\,000 samples for each temperature and method. This allows us to account for statistical uncertainties inherent in our stochastic approach. Furthermore, it enables us to base our assessment not on how closely the model matches the observed data, but rather on the model's saturation, defined as the point at which the model has visited enough samples to accurately represent the target distribution, and providing additional samples becomes redundant (this corresponds to the convergence criterion we intend to use in future applications). Reported values thus correspond to the average of these 10 estimates, with uncertainties obtained by calculating the standard deviation. The uncertainties on the combinations of PULSE estimates are obtained by error propagation. For Wang–Landau simulations, the initial modification factor is set to $f_0 = \exp(1)$, and the flatness criterion to $80\%$,  evaluated every 100 Monte Carlo iterations. For the Wolff simulations, we perform an initial equilibration stage of 200 Monte Carlo steps, followed by 800 additional equilibration steps and 1000 measurements.

\begin{figure}
    \centering
    \includegraphics[width=\linewidth]{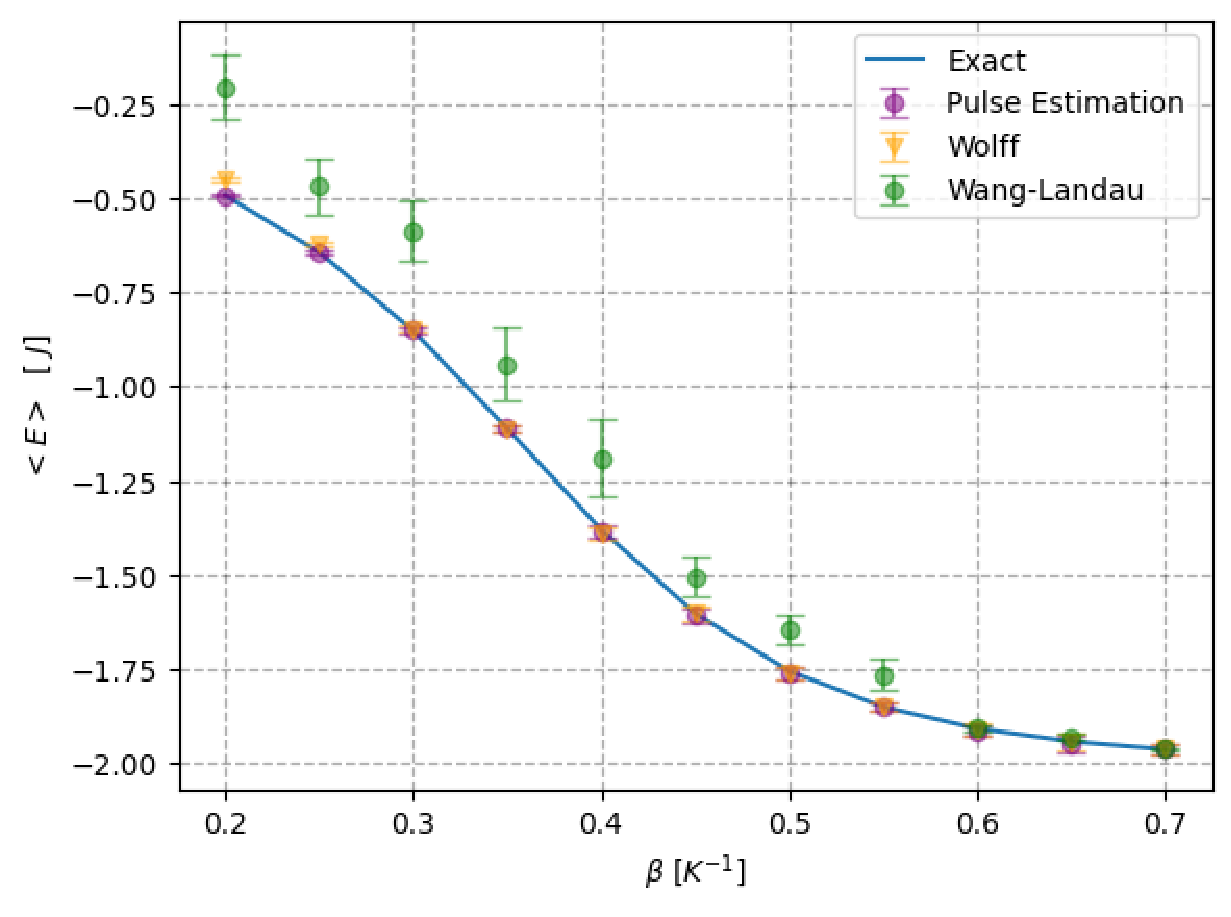}
    \caption{Comparison of sampling methods for the average energy of the 2D Ising system using PULSE, Wolff, and Wang-Landau algorithms.}
    \label{fig:wolffvPULSE_E}
\end{figure}

\begin{figure}
    \centering
    \includegraphics[width=\linewidth]{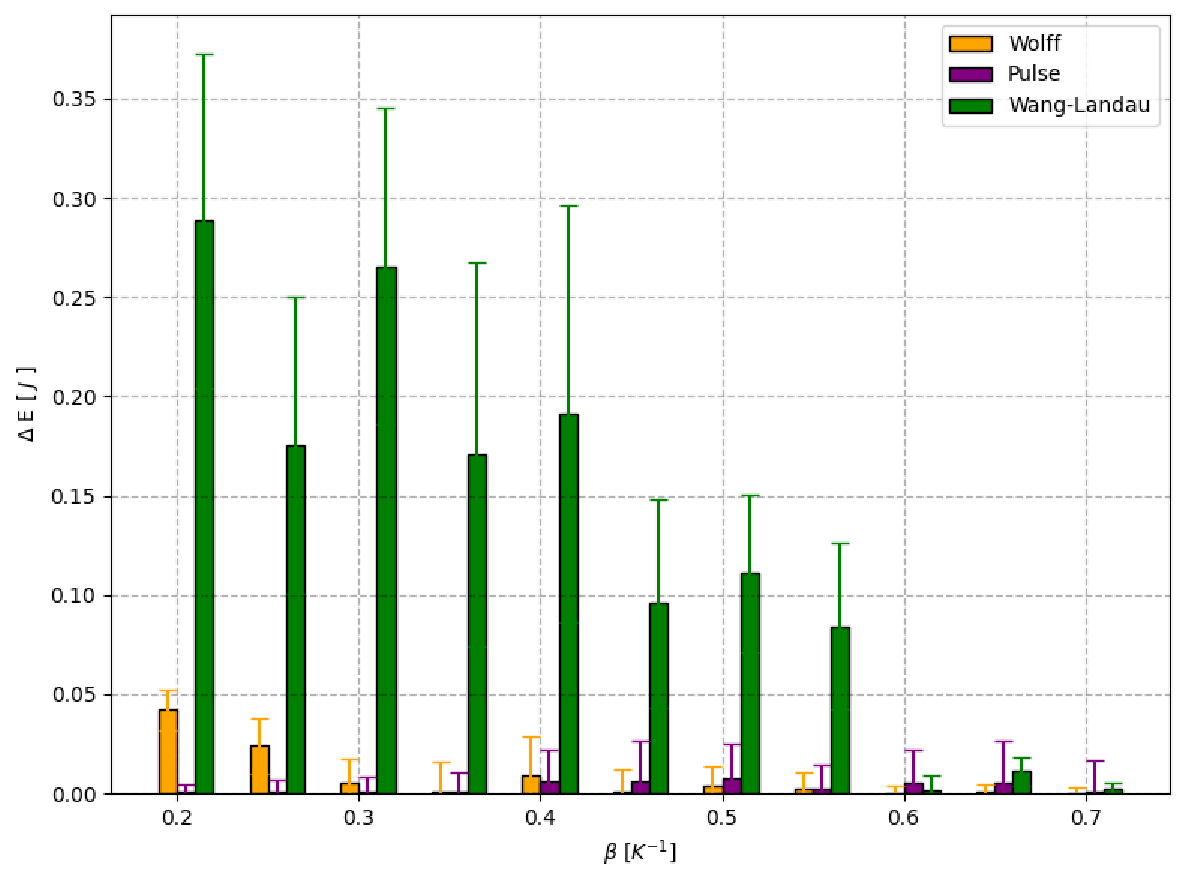}
    \caption{Comparison of errors in the sampling of the average energy using PULSE, Wolff, and Wang-Landau algorithms.}
    \label{fig:errwolffvPULSE_E}
\end{figure}

Figure \ref{fig:wolffvPULSE_E} shows the results of the three methods for the mean energy of the Ising system. 
All methods produce values close to the exact results, except for Wang-Landau at low $\beta$ (we observe for instance error exceeding 50\% at $\beta = 0.2$ K$^{-1}$).
Figure \ref{fig:errwolffvPULSE_E} shows the corresponding errors relative to the exact solution, along with uncertainty estimates. It should be noted that each method yields better results in different temperature regimes. Wolff's algorithm performs better at low temperatures ($\beta \gtrsim 0.4$ K$^{-1}$), because, as mentioned above, PULSE has difficulties in finding representative states in this temperature range. In contrast, Wolff's algorithm easily finds these states where the spins are all aligned thanks to its algorithm allowing updating aligned clusters all at ones. In the other temperature regime ($\beta \lesssim 0.4$ K$^{-1}$), characterized by disordered states, the opposite behaviour is observed: PULSE achieves a very high accuracy, whereas Wolff's error increases with temperature and with the number of significant Ising states in the energy calculation. Finally, Wang-Landau performs well at very low temperatures ($\beta \gtrsim 0.6$ K$^{-1}$), where it finds the representative states of the partition function easily, but very poorly at higher temperature. This behaviour is expected, as this method is not designed to estimate mean properties in the same way as standard Monte Carlo methods, and it thus needs significantly more samples to converge in the high-temperature regime. Indeed, since it explores configurations regardless of their energy, it requires more configurations to converge when there are more energy levels that are significant for the partition function.

\begin{figure}
    \centering
    \includegraphics[width=\linewidth]{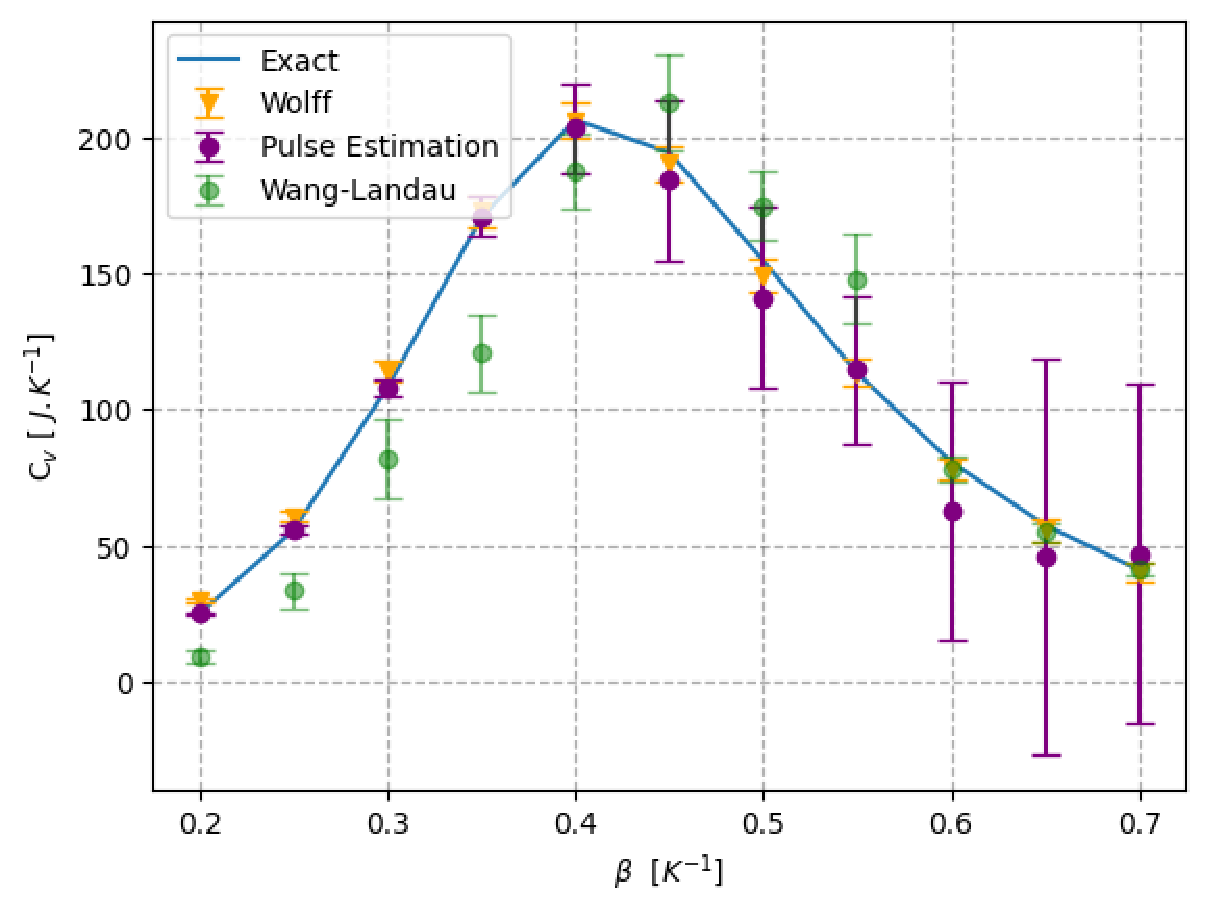}
    \caption{Comparison of sampling of mean specific heat by PULSE, Wolff and Wang-Landau with error bars.}
    \label{fig:wolffvPULSE_S}
\end{figure}

\begin{figure}
    \centering
    \includegraphics[width=\linewidth]{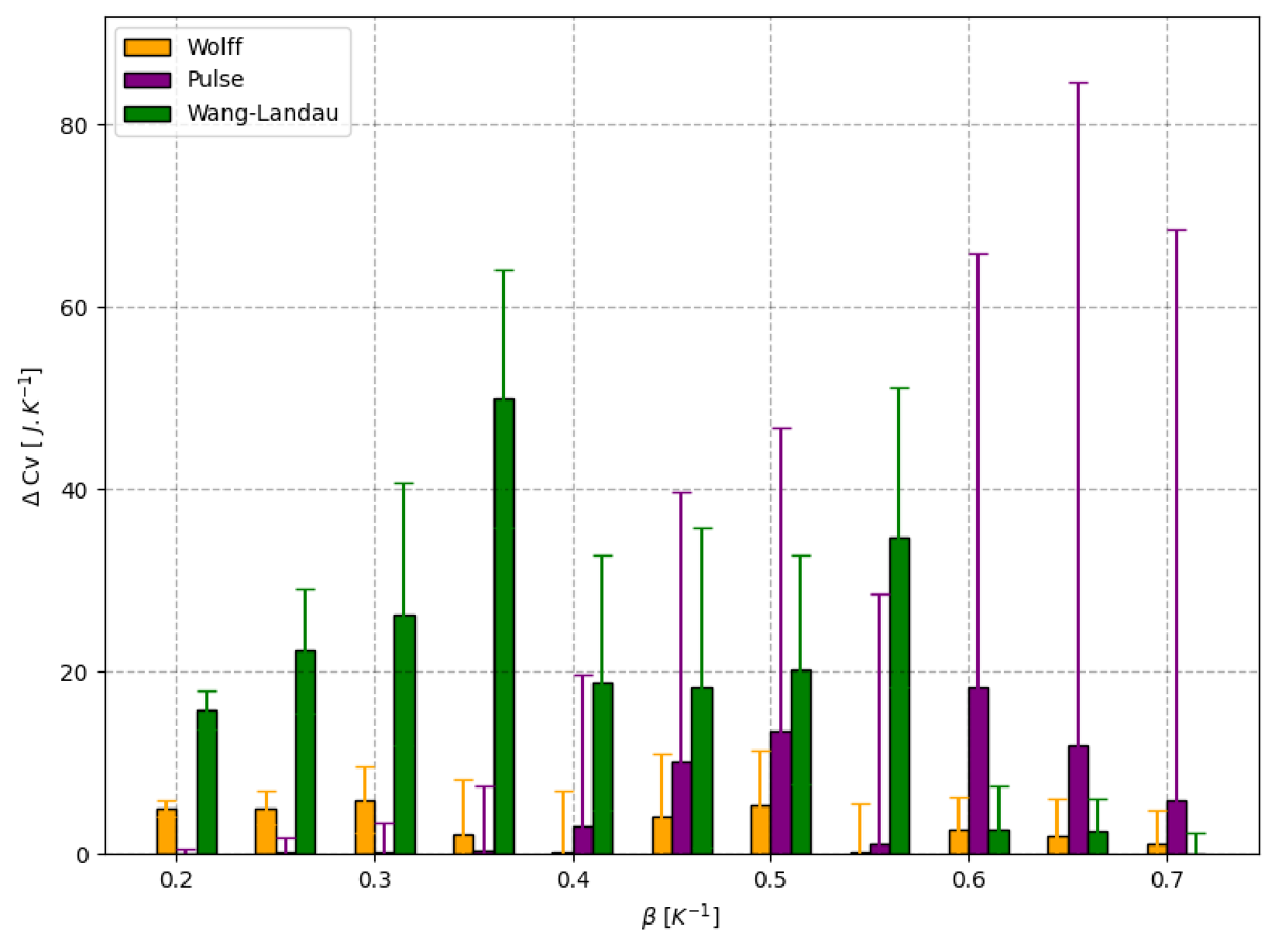}
    \caption{Comparison of errors in the sampling of the average specific heat by PULSE, Wolff, and Wang-Landau algorithms.}
    \label{fig:errwolffvPULSE_S}
\end{figure}

Finally, these energy estimates are used to reconstruct the target property, i.e., the specific heat at constant volume (cf. Eq. \ref{eq:spheat}). We show in Fig. \ref{fig:wolffvPULSE_S} the results of all methods, and in Fig. \ref{fig:errwolffvPULSE_S} the corresponding errors.

For this property, the previously observed trends are confirmed. Figure \ref{fig:wolffvPULSE_S} and \ref{fig:errwolffvPULSE_S} clearly show that each sampling method seems more suited to a specific temperature range. 
As the PULSE method struggles to find the few most significant states at low temperature, the deviations from the exact solution are slightly larger, and the error bars more significant in this temperature range. On the other hand, for small $\beta$ ($< 0.4$ K$^{-1}$), PULSE is in excellent agreement with the exact value, better than Wolff in this temperature range. Also, we find again that Wang-Landau performs well only at very low temperatures ($\beta \gtrsim 0.6$ K$^{-1}$).

The behaviour of our PULSE algorithm is therefore promising for future applications addressing the thermodynamic properties of disordered materials. With a relatively low number of samples, PULSE provides excellent estimates of the target property, outperforming the capabilities of the benchmark Monte Carlo methods in the high-temperature regime, while keeping the error in the low-temperature regime at acceptable levels. Most notably, it is usually the high-temperature range, where PULSE performs best, that is the most interesting for many applications, including nuclear ones. For this reason, conditions where very few configurations only are representative of the entire system (effectively tending towards a more ordered compound) is not a typical situation that this algorithm has been designed to treat. Achieving high accuracy is particularly important when evaluating combined physical quantities, for which errors in the underlying quantities can propagate and accumulate, as well as when the resulting quantities are subsequently used as inputs to models at larger scales. 
PULSE can therefore be more computationally efficient than the Monte Carlo methods considered here because it can achieve a given level of accuracy with substantially fewer samples, an advantage that becomes particularly significant when each energy evaluation is computationally costly, as will be the case in future applications to real materials.
The comparison with Wolff should therefore be understood primarily as a validation of the accuracy of PULSE, rather than as a claim that PULSE will outperform Wolff in every sampling scenario. Finally, contrary to most sampling methods currently available, a key advantage of PULSE is that it provides both a way to sample the partition function of the system and an estimate of it, which is particularly valuable for physical exploitation.

Finally, we can add a perspective for improving our model in light of these results. Our model operates with few assumptions about the configuration space to be explored, and appears to have difficulties sampling very specific cases, \textit{i.e.}, finding configurations with all spins aligned at low temperature. One idea for the future development of PULSE would be to add assumptions about the physical environment to be studied, if we have prior knowledge of it. For instance, in our previous work \cite{karcz2024targeting} the PULSE loss function was tweaked to take into account the nominal Pu concentration in the generated atomic configurations. 

\section{\label{sec:conclusions}Conclusions}

In this work, we extended the framework that allows the PULSE method to estimate thermodynamic properties of the 2D Ising system through optimized sampling of the configuration space. The method offers several computational advantages relevant to the study of disordered materials that will be targeted in future work:

\begin{itemize}
    \item[-] It estimates target properties with good accuracy from a small number of samples, and therefore at low computational cost. 
    
    \item[-] Its sampling cost is stable and essentially independent of system size, so that large configuration spaces can be treated without enlarging the model or increasing the number of samples. 

    \item[-] A single generative model performs both the sampling and the estimation and, unlike most conventional sampling methods, provides an explicit estimate of the partition function itself, which is valuable for the physical exploitation of the results.
\end{itemize}

We validated these properties on the two-dimensional Ising model, used as an exactly-solvable model system for a two-component disordered compound, and benchmarked the method against the Wolff and Wang-Landau algorithms. PULSE is particularly effective at sampling disordered states in the high-temperature regime, and less accurate in the low-temperature, highly ordered regime, where errors accumulate across the combined partition functions. The low-temperature regime however corresponds to an extreme case, dominated by a few ordered states, that is unlikely to be encountered in real disordered compounds.

Considering the excellent results on the Ising system, the PULSE method fully achieves its objective of providing an efficient sampling and evaluation of the properties of disordered systems, as we had already begun in our previous study \cite{karcz2024targeting}. This paves the way to the application of the method to the thermodynamic properties of real disordered compounds, such as mixed-oxide nuclear fuels or high-entropy alloys; with applications that could make use of our generative tool, for example to generate training datasets for ML interatomic potentials.

The extension of the PULSE approach to realistic materials nevertheless raises several important challenges. In particular, incorporating physically relevant prior knowledge into the learning procedure will be essential to make the approach computationally efficient and reliable. In realistic materials, the configuration space can be extremely large, and an unconstrained data-driven exploration may lead the neural network to sample regions that are physically irrelevant or of limited interest for the application under consideration. An appropriate representation of the physically relevant configuration space, informed by the specific characteristics and constraints of the material, will therefore be a key aspect of future developments.

In conclusion, PULSE represents a significant advance in addressing the properties of chemically disordered compounds, providing a computationally efficient way to sample their partition function while estimating target properties of interest for the technological application of these materials. In this respect, its upcoming application to thermodynamic properties is expected to substantially enhance the characterization of these compounds. 

\section*{Acknowledgments}
This work was performed using HPC resources from GENCI TGCC Joliot-Curie under allocation 2025-A0180906008.
It contributes to the CEA RTA/RCOMB project and received funding from the CEA CFR program.
The authors thank R. Santet for fruitful discussions about the mathematical framework.

\section*{Data availability statement}
The data that supports the findings of this study are available from the corresponding author upon reasonable request.

\section*{Author declarations}
\subsection*{Conflict of interest}
The authors have no conflicts to disclose.

\subsection*{Declaration of generative AI and AI-assisted technologies in the manuscript preparation process}
During the preparation of this work, the authors used the Mistral Medium 3.5 model, deployed in the "Maïa-CHAT" internal CEA tool, to refine the language and improve the readability of the manuscript. The authors reviewed and edited the output as needed and take full responsibility for the content of the published article.

\section*{Appendix}
\appendix
\section{Probability definition}

It should be noted that the formulation of the partition function for an observable $O$ is as follows:

\begin{equation}
\begin{split}
    Z_{O} (T) = \sum_{x \in X} O(x) \exp\left(\frac{-E(x)}{k_{B}T}\right)  \\
    \text{By factorising the exponential, we can then write:} \\
    Z_{O} (T) = \sum_{x \in X} \exp\left(\frac{-E(x)}{k_{B}T} + \ln(O(x)) \right) \; . \\
\end{split}
\end{equation}

Eq. 3 is therefore obtained following this formulation to define the probability of finding a state $x$ with the value $O$:

\begin{equation}
    P_{O}(x) = \frac{\exp\left(\frac{-E(x)}{k_{B}T} + \ln(O(x)) \right)}{Z_{O}(T)} \; .
\end{equation}

\section{Definition of joint distributions}

Following previous works \cite{karcz2024targeting, cantwell2022approximate} function $R_{\phi}(x|y)$ is defined as a product distribution, with $\phi$ being a neural network:

\begin{equation}
    R_{\phi}(x|y) = \prod_i \frac{e^{I(x^{(i)}) \, \phi^{(i)}(y)}} {e^{\phi^{(i)}(y)} + e^{-\phi^{(i)}(y)}}  \; ,
\end{equation}

with $I$ an indicator function defined as follows for the Ising system:

\begin{equation}
    I(x^{(i)}) = \left\{
    \begin{array}{l}
    \quad 1 \quad \text{if $x^{(i)}$ represents a spin up,} \\
    \, -1 \quad \text{if $x^{(i)}$ represents a spin down.}  \\
    \end{array}
    \right.
\end{equation}

$Q_{\theta}(y|x)$ follows the same logic but reversed, with $\theta$ being a neural network:

\begin{equation}
    Q_{\theta}(y|x) = \prod_i \frac{e^{y^{(i)}\theta^{(i)}(x)}} {e^{\theta^{(i)}(x)} + e^{-\theta^{(i)}(x)}}  \; .
\end{equation}

\printcredits

\bibliographystyle{unsrt}
\bibliography{P40_PULSE_Ising}

@article{karcz2024targeting,
  title={AI-driven data-efficient estimation of partition functions in disordered materials},
  author={Karcz, Maciej J and Messina, Luca and Kawasaki, Eiji and Bourasseau, Emeric},
  journal={Scientific Reports},
  volume={16},
  pages={14568},
  year={2026},
  publisher={Nature Publishing Group UK London}
}

@article{wolff1989collective,
  title={Collective Monte Carlo updating for spin systems},
  author={Wolff, Ulli},
  journal={Physical Review Letters},
  volume={62},
  number={4},
  pages={361},
  year={1989},
  publisher={APS}
}

@article{cantwell2022approximate,
  title={Approximate sampling and estimation of partition functions using neural networks},
  author={Cantwell, George T},
  journal={arXiv preprint arXiv:2209.10423},
  year={2022}
}

@article{d2020learning,
  title={Learning the Ising model with generative neural networks},
  author={D'Angelo, Francesco and B{\"o}ttcher, Lucas},
  journal={Physical Review Research},
  volume={2},
  number={2},
  pages={023266},
  year={2020},
  publisher={APS}
}

@article{kingma2013auto,
  title={Auto-encoding variational bayes},
  author={Kingma, Diederik P and Welling, Max},
  journal={arXiv preprint arXiv:1312.6114},
  year={2013}
}

@article{kingma2014adam,
  title={Adam: A method for stochastic optimization},
  author={Kingma, Diederik P and Ba, Jimmy},
  journal={arXiv preprint arXiv:1412.6980},
  year={2014}
}

@article{landau1976finite,
  title={Finite-size behavior of the Ising square lattice},
  author={Landau, DP},
  journal={Physical Review B},
  volume={13},
  number={7},
  pages={2997},
  year={1976},
  publisher={APS}
}

@article{alexandrou2020critical,
  title={The critical temperature of the 2D-Ising model through deep learning autoencoders},
  author={Alexandrou, Constantia and Athenodorou, Andreas and Chrysostomou, Charalambos and Paul, Srijit},
  journal={The European Physical Journal B},
  volume={93},
  number={12},
  pages={226},
  year={2020},
  publisher={Springer}
}

@article{zhang2021ising,
  title={Ising spin configurations with the deep learning method},
  author={Zhang, Yihang},
  journal={Journal of Physics Communications},
  volume={5},
  number={1},
  pages={015006},
  year={2021},
  publisher={IOP Publishing}
}

@article{onsager1944crystal,
  title={Crystal statistics. I. A two-dimensional model with an order-disorder transition},
  author={Onsager, Lars},
  journal={Physical review},
  volume={65},
  number={3-4},
  pages={117},
  year={1944},
  publisher={APS}
}

@article{lecun2015deep,
  title={Deep learning},
  author={LeCun, Yann and Bengio, Yoshua and Hinton, Geoffrey},
  journal={nature},
  volume={521},
  number={7553},
  pages={436--444},
  year={2015},
  publisher={Nature Publishing Group UK London}
}

@article{kullback1951information,
  title={On information and sufficiency},
  author={Kullback, Solomon and Leibler, Richard A},
  journal={The annals of mathematical statistics},
  volume={22},
  number={1},
  pages={79--86},
  year={1951},
  publisher={JSTOR}
}

@article{OSTOVARIMOGHADDAM2021105504,
title = {Does the pathway for development of next generation nuclear materials straightly go through high-entropy materials?},
journal = {International Journal of Refractory Metals and Hard Materials},
volume = {97},
pages = {105504},
year = {2021},
issn = {0263-4368},
doi = {https://doi.org/10.1016/j.ijrmhm.2021.105504},
url = {},
author = {Ahmad {Ostovari Moghaddam} and Andreu Cabot and Evgeny A. Trofimov}
}

@Inbook{Gao2016,
author="Gao, Michael C.
and Niu, Changning
and Jiang, Chao
and Irving, Douglas L.",
title="Applications of Special Quasi-random Structures to High-Entropy Alloys",
bookTitle="High-Entropy Alloys: Fundamentals and Applications",
year="2016",
publisher="Springer International Publishing",
address="Cham",
pages="333--368",}

@article{wetzel2017unsupervised,
  title={Unsupervised learning of phase transitions: From principal component analysis to variational autoencoders},
  author={Wetzel, Sebastian J},
  journal={Physical Review E},
  volume={96},
  number={2},
  pages={022140},
  year={2017},
  publisher={APS}
}

@article{hu2017discovering,
  title={Discovering phases, phase transitions, and crossovers through unsupervised machine learning: A critical examination},
  author={Hu, Wenjian and Singh, Rajiv RP and Scalettar, Richard T},
  journal={Physical Review E},
  volume={95},
  number={6},
  pages={062122},
  year={2017},
  publisher={APS}
}

@article{torlai2016learning,
  title={Learning thermodynamics with Boltzmann machines},
  author={Torlai, Giacomo and Melko, Roger G},
  journal={Physical Review B},
  volume={94},
  number={16},
  pages={165134},
  year={2016},
  publisher={APS}
}

@Article{e15104504,
AUTHOR = {Gao, Michael C. and Alman, David E.},
TITLE = {Searching for Next Single-Phase High-Entropy Alloy Compositions},
JOURNAL = {Entropy},
VOLUME = {15},
YEAR = {2013},
NUMBER = {10},
PAGES = {4504--4519},
ISSN = {1099-4300},
DOI = {10.3390/e15104504}
}

@article{TAN2023100114,
title = {Multiscale modelling of irradiation damage behavior in high entropy alloys},
journal = {Advanced Powder Materials},
volume = {2},
number = {3},
pages = {100114},
year = {2023},
issn = {2772-834X},
doi = {https://doi.org/10.1016/j.apmate.2023.100114},
author = {Fusheng Tan and Li Li and Jia Li and Bin Liu and Peter K. Liaw and Qihong Fang},
}

@article{walker2020deep,
  title={Deep learning on the 2-dimensional Ising model to extract the crossover region with a variational autoencoder},
  author={Walker, Nicholas and Tam, Ka-Ming and Jarrell, Mark},
  journal={Scientific reports},
  volume={10},
  number={1},
  pages={13047},
  year={2020},
  publisher={Nature Publishing Group UK London}
}

@article{yevick2022variational,
  title={Variational autoencoder analysis of Ising model statistical distributions and phase transitions},
  author={Yevick, David},
  journal={The European Physical Journal B},
  volume={95},
  number={3},
  pages={56},
  year={2022},
  publisher={Springer}
}

@article{pickering2021high,
  title={High-entropy alloys for advanced nuclear applications},
  author={Pickering, Ed J and Carruthers, Alexander W and Barron, Paul J and Middleburgh, Simon C and Armstrong, David EJ and Gandy, Amy S},
  journal={Entropy},
  volume={23},
  number={1},
  pages={98},
  year={2021},
  publisher={MDPI}
}

@incollection{ZINKLE2017569,
title = {16 - Advanced irradiation-resistant materials for Generation IV nuclear reactors},
editor = {Pascal Yvon},
booktitle = {Structural Materials for Generation IV Nuclear Reactors},
publisher = {Woodhead Publishing},
pages = {569-594},
year = {2017},
isbn = {978-0-08-100906-2},
doi = {https://doi.org/10.1016/B978-0-08-100906-2.00016-1},
author = {S.J. Zinkle}
}

@article{case2016convergence,
  title={Convergence properties of crystal structure prediction by quasi-random sampling},
  author={Case, David H and Campbell, Josh E and Bygrave, Peter J and Day, Graeme M},
  journal={Journal of chemical theory and computation},
  volume={12},
  number={2},
  pages={910--924},
  year={2016},
  publisher={ACS Publications}
}

@article{wei2018metastability,
  title={Metastability in high-entropy alloys: A review},
  author={Wei, Shaolou and He, Feng and Tasan, Cemal Cem},
  journal={Journal of Materials Research},
  volume={33},
  number={19},
  pages={2924--2937},
  year={2018},
  publisher={Cambridge University Press}
}

@article{tsai2014high,
  title={High-entropy alloys: a critical review},
  author={Tsai, Ming-Hung and Yeh, Jien-Wei},
  journal={Materials Research Letters},
  volume={2},
  number={3},
  pages={107--123},
  year={2014},
  publisher={Taylor \& Francis}
}

@article{liu2023machine,
  title={Machine learning for high-entropy alloys: Progress, challenges and opportunities},
  author={Liu, Xianglin and Zhang, Jiaxin and Pei, Zongrui},
  journal={Progress in Materials Science},
  volume={131},
  pages={101018},
  year={2023},
  publisher={Elsevier}
}

@article{alamino2024explaining,
  title={Explaining the Machine Learning Solution of the Ising Model},
  author={Alamino, Roberto C},
  journal={arXiv preprint arXiv:2402.11701},
  year={2024}
}

@article{zhang2008solid,
  title={Solid-solution phase formation rules for multi-component alloys},
  author={Zhang, Yong and Zhou, Yun Jun and Lin, Jun Pin and Chen, Guo Liang and Liaw, Peter K},
  journal={Advanced engineering materials},
  volume={10},
  number={6},
  pages={534--538},
  year={2008},
  publisher={Wiley Online Library}
}

@article{wei1990electronic,
  title={Electronic properties of random alloys: Special quasirandom structures},
  author={Wei, S-H and Ferreira, LG and Bernard, James E and Zunger, Alex},
  journal={Physical Review B},
  volume={42},
  number={15},
  pages={9622},
  year={1990},
  publisher={APS}
}

@article{shin2006thermodynamic,
  title={Thermodynamic properties of binary hcp solution phases from special quasirandom structures},
  author={Shin, Dongwon and Arr{\'o}yave, Raymundo and Liu, Zi-Kui and Van de Walle, Axel},
  journal={Physical Review B—Condensed Matter and Materials Physics},
  volume={74},
  number={2},
  pages={024204},
  year={2006},
  publisher={APS}
}

@article{vigier2015structural,
  title={Structural investigation of (U0. 7Pu0. 3) O2-x mixed oxides},
  author={Vigier, Jean-Francois and Martin, Philippe M and Martel, Laura and Prieur, Damien and Scheinost, Andreas C and Somers, Joseph},
  journal={Inorganic chemistry},
  volume={54},
  number={11},
  pages={5358--5365},
  year={2015},
  publisher={ACS Publications}
}

@article{jang2016categorical,
  title={Categorical reparameterization with gumbel-softmax},
  author={Jang, Eric and Gu, Shixiang and Poole, Ben},
  journal={arXiv preprint arXiv:1611.01144},
  year={2016}
}

@article{wang2001determining,
  title={Determining the density of states for classical statistical models: A random walk algorithm to produce a flat histogram},
  author={Wang, Fugao and Landau, DP},
  journal={Physical Review E},
  volume={64},
  number={5},
  pages={056101},
  year={2001},
  publisher={APS}
}

@article{lagache2001prediction,
  title={Prediction of thermodynamic derivative properties of fluids by Monte Carlo simulation},
  author={Lagache, M and Ungerer, Ph and Boutin, A and Fuchs, AH},
  journal={Physical Chemistry Chemical Physics},
  volume={3},
  number={19},
  pages={4333--4339},
  year={2001},
  publisher={Royal Society of Chemistry}
}

@article{TAKOUKAMTAKOUNDJOU2020152125,
title = {Study of thermodynamic properties of U1-yPuyO2 MOX fuel using classical molecular Monte Carlo simulations},
journal = {Journal of Nuclear Materials},
volume = {534},
pages = {152125},
year = {2020},
issn = {0022-3115},
doi = {https://doi.org/10.1016/j.jnucmat.2020.152125},
author = {Cyrille Takoukam-Takoundjou and Emeric Bourasseau and Véronique Lachet},
}

@article{murtazaev2015critical,
  title={Critical properties of the two-dimensional Ising model on a square lattice with competing interactions},
  author={Murtazaev, AK and Ramazanov, MK and Badiev, MK},
  journal={Physica B: Condensed Matter},
  volume={476},
  pages={1--5},
  year={2015},
  publisher={Elsevier}
}

@article{albano2003corner,
  title={Corner wetting in the two-dimensional Ising model: Monte Carlo results},
  author={Albano, EV and De Virgiliis, A and M{\"u}ller, M and Binder, K},
  journal={Journal of Physics: Condensed Matter},
  volume={15},
  number={3},
  pages={333--345},
  year={2003}
}

@article{shchur2001critical,
  title={Critical amplitude ratio of the susceptibility in the random-site two-dimensional Ising model},
  author={Shchur, Lev N and Vasilyev, Oleg A},
  journal={Physical Review E},
  volume={65},
  number={1},
  pages={016107},
  year={2001},
  publisher={APS}
}

%



\end{document}